# In-situ high energy synchrotron X-ray diffraction investigation of the $MgB_2$ phase formation and $MgB_2$ tapes sintering

Maurizio Vignolo[1,*], Gennaro Romano[1], Davide Nardelli[2], Emilio Bellingeri[1], Alberto Martinelli[1], Alexey Bitchkov[3], Cristina Bernini[1], Andrea Malagoli[1], Valeria Braccini[1] and Carlo Ferdeghini[1]

[1] CNR-SPIN Genova, C.so Perrone 24, 16152 Genova-Italy
[2] Former CNR- SPIN, now ASG Superconductors
[3] ESRF Grenoble, France

*Corresponding author. Tel. +39-010-6598789; e-mail: maurizio.vignolo@spin.cnr.it

Abstract: In the present paper we report an *in-situ* high-energy X-ray diffraction analysis of $MgB_2$ tapes during the preparation process. The experiment was performed in a specifically designed furnace working in reducing atmosphere, compatible with the Laue diffraction condition. The $MgB_2$ synthesis was realized starting from $MgH_2$ and amorphous B in powder form as precursors, varying reaction temperature and testing different cooling processes. We analyzed both the $MgB_2$ synthesis and the sintering process of tapes prepared with these powders. Phase evolution, micro and crystallographic structure were monitored during the different thermal treatments. Among the main results we observed the formation of $MgB_2$ at an extraordinary low temperature (300°C), probably as a result of a solid-state reaction between $MgH_2$ and B. Furthermore, we studied the dependence of the micro-structure upon the thermal treatment and its effect on the critical current performance of the superconducting tapes.

## Introduction

$MgB_2$ conductors are now fabricated in long lengths and commercialized for some particular applications such as low field MRI systems: if on one side the technological advancement has been significant, still a strong research effort is required. In fact, in order to be able to widen the application range of $MgB_2$ conductors, a significant enhancement of $J_c$ is needed, which came to a standstill and does not look as a trivial task at the moment [1]. If it has been shown that $H_{c2}$ can be as high as 60 T [2] in thin films, although numerous attempts have been done it stays below 35 T in polycrystals [3] while at the same time $J_c$ looks limited. The research should be focused then on understanding how it is possible to efficiently add different pinning centres and / or new pinning mechanisms.

So far, the most successful results in terms of $J_c$ reached in polycrystals have been obtained via doping with carbon under different forms [4], or combining carbon doping and mechanical alloying [5]. Concerning the preparation of *ex-situ* Powder-In-Tube (P.I.T.) $MgB_2$ conductors for applications, we obtained good $J_c$ values by applying high-energy ball milling [6] and adding carbon [7] or by lowering the synthesis temperature [8].

Given that the main pinning mechanism in $MgB_2$ is due to grain boundaries [9], the reduced grain size obtained through these two ways helped significantly in improving the critical current in-field behaviour.

In a previous work, the influence of the synthesis temperature ($T_S$) and processing atmosphere during the cooling process on the $MgB_2$ grain size were investigated and correlated to *ex-situ* P.I.T. tape performances [10]: in this work we will closely examine these topics, with particular focus on the formation of the phase. In order to clarify all the aspects related to the thermal treatments in the tape realization we performed high-energy X-ray diffraction (HEXRD) during both the synthesis of $MgB_2$ and the tape sintering process of an *ex-situ* $MgB_2$ tape. To reduce the probability of uncontrolled oxidation during the synthesis we used, as suggested in the literature [11-13], $MgH_2$ instead of metallic Mg as a precursor.

The use of the *in-situ* analysis like neutron or HEXRD is a powerful tool [14-17] for studying the synthesis and the sintering processes involved during the preparation of $MgB_2$ tapes. These techniques have been already applied to investigate the stress induced during the deformation process and the subsequent re-crystallization as a function of the sintering temperature [14]; as a result the optimal sintering temperature was identified to be 920 °C because, at higher temperatures, Ni reacts with magnesium diboride forming a layer of $Ni_{2.5}MgB_2$. $MgB_2$ formation was previously studied by HEXRD. In particular it was observed that the reaction time is decreased from 100 up to 2 minutes if the temperature is increased from 670 up to 900°C [15]. The formation of MgO and $MgB_2$ was observed at 450°C and 575°C, respectively, investigating the synthesis of $MgB_2$ applying the *in-situ* P.I.T. method [16]. Finally the $MgB_2$ formation in the Mg-B-N system at high pressure (5.5GPa) and temperature (1900 K) was studied, revealing that the $MgB_2$ formation starts with a solid-state reaction [17].

Summarizing, in this paper we studied the $MgB_2$ synthesis at various temperatures by *in-situ* HEXRD, using $MgH_2$ and amorphous B as precursors and analysed the effect of different atmosphere during the cooling process. Moreover, the sintering process of tapes prepared with these powders was also examined relating micro and crystallographic structure with the critical current performance of superconducting tapes.

**Experimental details**

HEXRD analysis was carried out at the ID11 beamline of the European Synchrotron Radiation Facility (ESRF) in Grenoble. The set up of the experiment, similar to that reported in [18,19], is sketched in figure 1. A tubular furnace has been placed on the high energy X-ray beam path (88 KeV, $\lambda$ = 0.1370 Å) for transmission measurements. The quartz tube, with a inner diameter of 60 mm, was sealed with Kapton windows on appropriate PVC caps at both ends. The caps were equipped with a gas inlet or outlet, respectively, to maintain a controlled Ar-5% $H_2$ atmosphere during each heat treatment. Two perforated discs of refractory material were mounted in order to limit irradiation and maintain higher homogeneity into the central zone of the furnace, where the sample is placed. The samples were aligned throughout an optical system.

Laue diffraction patterns were acquired by a CCD camera (10 patterns/min) and were integrated to obtain the diffracted intensity as a function of the diffracting angle ($I$-$2\theta$ diffraction patterns). Rietveld refinement [20] was carried out using FULLPROF software [21]. By means of a $LaB_6$ polycrystalline standard an instrumental resolution file was obtained and applied during refinements. In this way the variation of the cell parameters as a function of the reaction time/temperature was investigated as well as the phase evolution.

The apparatus above described and the HEXRD measurement technique were used to study the various steps of the process preparation of an *ex-situ* $MgB_2$ tape [22].
The *ex-situ* P.I.T. process is a very flexible method to tailor the properties of superconducting powder and can be divided in the following steps:

i) $MgB_2$ synthesis from precursors,
ii) possible powder treatments (mechanical, thermal, oxidation),
iii) filling of the tube with powders followed by tube sealing,
iv) mechanical deformation to obtain tapes, wires or cables,
v) final sintering process.

In this experiment we studied the thermal treatments performed during the *ex-situ* P.I.T. process, i.e. the steps i), ii) and v). We analyzed this last step both for the powders prepared in inert atmosphere and for the powders exposed to oxygen in the step ii). In fact it is during these heat treatments that it is possible to tune the $MgB_2$ properties from the chemical point of view.
Regarding the $MgB_2$ synthesis (i), HEXRD data were acquired during two reactions at 760°C (*LT*) and 910°C (*HT*), respectively. We then studied the oxidation process during cooling (ii). The deformation process applied was the "standard" one [22] optimized for maximizing the critical current for our powders. Summarizing, 3 kinds of tapes were analyzed, as shown in Table 1.

| *Sample name* | *Powder reaction temperature($T_s$)* [°C] | *Power oxidized during cooling process* |
|---|---|---|
| LTIA | 760 | No |
| LTOA | 760 | Yes |
| HTIA | 910 | No |

Table 1. List of samples analyzed through HEXRD.

In the following we will refer to the different typologies with the sample names of the table 1.

## Results and discussion

### *Powder synthesis and phase evolution*

As suggested in the literature [11-14] and with the aim to reduce the possible uncontrolled oxidation, we used $MgH_2$ instead of metallic Mg as a precursor for synthesizing $MgB_2$ involved into the *ex-situ* process. In this way a possible oxygen source is eliminated since it is well known that metallic Mg is easy oxidised by atmospheric $O_2$, both during storage and manipulation. Furthermore, the formation of $H_2$ by $MgH_2$ decomposition produces a reducing atmosphere which hinders oxygen contamination.

The reactions involved in the $MgB_2$ synthesis can be summarized by the following main reactions (1)-(3):

$$MgH_{2(s)} = Mg_{(s)} + H_{2(g)} \text{ at } T = 450°C \ (\Delta_r H° > 0, \Delta_f G° = -80 KJ/mol \text{ for } MgH_2) \quad (1)$$
$$Mg_{(s)} = Mg_{(l)} \text{ at } T = 650°C \quad (2)$$
$$Mg_{(l)} + 2B_{(s)} = MgB_{2(s)} \quad (3)$$

In our experiment, the heating rate for the synthesis reaction was 10°C/min and samples were kept for one hour at the regime temperature ($T_S$); then they were cooled at 10°C/min down to 500°C and then freely down to room temperature by opening the upper part of the split furnace. As earlier mentioned, in this experiment we tested two different $T_S$: 760 and 910°C.

In order to investigate the synthesis reaction with the same experimental set-up used at SPIN laboratories we tried to study the $MgB_2$ powders synthesis into very thin walls – i.e. 100 μm - Fe crucibles. Unfortunately, many Fe peaks lie on the same angular positions of crucial $MgB_2$ and MgO peaks, then we also analyzed some pellets of pressed precursor powders without any sheath material. The pellets, with 1 cm diameter and height, were prepared mixing the precursor and pressing them with a 2 tons force. Both syntheses were carried out under an inert atmosphere (*IA*) at two different synthesis temperatures: 760°C (low temperature inert atmosphere or *LTIA* sample) and 910°C (high temperature inert atmosphere or *HTIA* sample).

Great care was employed in order to reproduce the standard preparation conditions applied at the SPIN laboratories; however we underline how during the experiment at ESRF it was necessary to carry out parts of the process in air and not in inert atmosphere glove box: in some respects this could make a difference. Moreover, it should be noted that the actual temperature of the pellet can be significantly different from that of the furnace. In figure 2a we show the furnace thermal profile compared with the pellet temperature measured by a thermocouple in direct contact with the powders in a similar system of figure 1. Here, the heating rate was 10°C/min, up to the reaction temperature, the same conditions we used for the experiment. At low temperatures (time) the two temperatures significantly differ; the difference between the furnace and pellet temperature tends to decrease at higher temperatures (time) but still remains appreciable. This is the reason why in the following figures we prefer to put in the x-axis the time and not directly the temperature. The estimated temperature is reported, for clarity, in the upper x-axis only as a reference. Looking in detail figure 2 an endothermic peak is observed in the temperature range between 450° and 550°C in both curves, due to the $MgH_2$ decomposition (figure 2b), as described later. A small exothermic peak is present around 700°C (figure 2c) in both thermal profiles (up to 760°C and up to 910°C): this peak results from the superposition of an endothermic peak, originated by Mg melting, and a stronger exothermic peak, resulting from the reaction between Mg and B. This thermal effect is indicative of the fact that as soon as Mg melts, it spreads throughout B powders and reacts with them forming $MgB_2$.

Figure 3 shows the phase evolution as volume percentage versus time for $MgB_2$ synthesis at 760°C as obtained from Rietveld refinement of HEXRD data. Metallic Mg (full triangle) is already present (7%) at the beginning of the reaction. This is probably due to an incomplete hydrogenation reaction or a decomposition reaction of $MgH_2$ during its storage. The formation of $MgB_2$ (full square) takes place at very low temperature (less than 300°C), around 30 minutes after the beginning of the thermal treatment. This exceptionally low temperature formation is probably due to a solid-solid reaction between $MgH_2$ and B: indeed the $MgH_2$ amount slightly decreases after 30 min (full circle), whereas the metallic Mg remains constant.

In the light of these results we must consider an another reaction involved in the $MgB_2$ synthesis, when $MgH_2$ is employed, which must be temporally placed before the step 1) and can be represented by the following equation:

$$MgH_{2(s)} + 2B_{(s)} = MgB_{2(s)} + H_{2(g)}$$

The solid state reaction between Mg and B was already reported in previous papers [23-26] but located at about 500°C. Furthermore, in those cases, metallic Mg was used instead of $MgH_2$. This lower reaction temperature may be due to a catalytic effect of $H_2$, even if this evidence has not been verified yet. The $MgB_2$ content exhibits a slight increase as the decomposition of $MgH_2$ takes place, in the temperature range 400-500°C, confirming what reported in [27]. At higher temperatures (times) only metallic Mg is present. The $MgB_2$ amount undergoes a dramatic increase when the temperature reaches the melting point of Mg (650°C) and for further 10 minutes. When the oven temperature reaches 760°C (after 76 min), the formation of $MgB_2$ is complete (97% volume). The remaining 3% is represented by MgO whose formation starts at the Mg melting point. A small amount of oxygen is always present, probably due to the refractory material used as shield in thermal insulation or physically absorbed during the pellet making-off.

The phase evolution analysis is in qualitative agreement with thermal analysis obtained during the low and high synthesis temperature in similar furnace at SPIN laboratory (figure 2). The synthesis reaction of $MgB_2$ described by equation (3) is complete in the temperature range 700-780°C. We did not record any significant difference in the phase evolution when the process has been carried out up to 910°C for *HTIA* sample except for a slightly reduced $MgB_2$ formation time due to the higher temperature. Differences, however, are present in the microstructure: in figure 4 we show the granulometry distribution for the two $MgB_2$ powders fitted with a Lorentzian function. Such distribution was obtained from a statistics on about 700 particles taken from several points of the two samples with a Scanning Electron Microscopy in order to estimate the average diameter of the powders. As it is clearly shown, the granulometry distribution is more homogeneous and has a lower maximum (0.3 μm vs. 1.4 μm) for the powders synthesized at low temperature, as a natural consequence of the different grain growth at different temperatures. Other microstructural differences are also evidenced during the subsequent sintering process of the tape as described in the next paragraph.

***Tape sintering***

The importance of the final sintering heating treatment is based on the necessity of recovering the lattice strain induced by the prolonged cold working procedure applied in the P.I.T. process and improving the grain connectivity and the $T_c$ as well (the cold deformation working alone can lower $T_c$ down to 32 K [22]). We therefore investigated the sintering process in inert atmosphere by *in-situ* HEXRD analysis to see its influence on the structural features.

Figure 5 shows the *a* and *c* cell parameters evolution during the sintering process, obtained by refinement of the HEXRD data. The upper panels refer to the tapes realized with *LT* powders and lower panels to the ones realized with *HT* powders. The *a* and *c* parameters for *HT* samples closely follow the temperature profile, differently from what is observed for *LT* where a strong rearrangement of the cell edges is observed in the temperature region 750-900 °C. This effect is due to a greater stress recovery in the sample synthesized at low temperatures whereas the *HT* sample has a more relaxed structure parameters already after synthesis.

In Figure 6 we show the broadening of the (100) peak during the sintering for the *LTIA* and *HTIA* samples. In both cases a structure relaxation trend is evident which ends with a higher value for the *LTIA* sample. This means a lower stress recovery and/or a reduced crystallite size for the sample *LTIA*. This trend is in good agreement with the data previously published in [14], where it was established that the cold working brings on a progressive peak broadening indicating the loss of crystallinity, followed by a recovery of crystallinity during the final heat treatment when the peaks became narrower again. It is interesting to note how the *LTIA* sample, at the end of the process, shows a more stressed structure and/or smaller grains than *HTIA*.

Figure 7 shows $J_c$ (from [6]) for two wires synthesized at low and high temperature in inert atmosphere. To correlate these data with those obtained from the HEXRD study we can speculate that the increased $J_c$ corresponding to the *LTIA* sample is due to a reduced granulometry which leads to an enhanced connectivity at all fields; furthermore, a higher contribute at high magnetic fields can be given by a bigger stress content.

## *Oxidation effects*

Finally, we studied the effect of oxidizing atmosphere during the cool down of the powders during the step ii) of the process. For this purpose we investigated by HEXRD tapes synthesized at low temperature in inert (*IA*) and oxidizing (*OA*) atmosphere during the cooling process (i.e. for T < 400°C).

In reference [8] we reported that MgO is present as a thin layer covering the $MgB_2$ grains: this layer hinders the grain growth during the sintering process. The same MgO layer (with a thickness comparable to the $MgB_2$ coherence length) can act as pinning centres enhancing the in-field $J_c$ performance of the tape.

To confirm the grain growth inhibition among *IA* and *OA* samples we have estimated the lattice rearrangement during the sintering process through the broadening of the full width at half maximum (FWHM) of the 100 peak because it is more sensitive to the sintering process from a structural arrangement point of view. The evolution with time of the FWHM of the 100 peak for the sample *LTOA* is reported in the same figure 6. From the plot it is clearly visible a higher peak broadening for the *LTOA* sample after sintering with respect to *LTIA,* whereas both the *LT* samples show the same re-crystallization trend during the whole process.

Figure 8 shows the comparison of the HEXRD patterns around the angular position of the 110 reflection of the $MgB_2$ phase of samples *LTOA* and *LTIA* after the sintering process. It is clearly visible that the MgO amount is higher in *OA* sample. This is consistent with the different critical current behaviour shown in figure 9, where we report the data of reference [10] for two samples realized as *LTOA* and *LTIA*.

We note how the best $J_c$ shown at low field is measured in the sample with the smallest MgO content (*LTIA*) and the best $J_c$ behaviour in high fields belong to the *LTOA* sample, which shows the highest MgO content and the more stressed structure. In reference [10] we showed how the grain boundary pinning is the main pinning mechanism in these powders namely when the oxygen introduction inhibits the grain growth. This is in agreement with the data of figure 6 where the peak broadening clearly increases after sintering for the oxidized sample. In addition the samples *LTIA* and *LTOA* presented the same critical temperature and the same critical field but considerable difference in the average grain size after sintering being in the case of *LTIA* nearly double [8]. The MgO at the grain boundaries decreases the connectivity (as resulted from the Rowell analysis of $\Delta\rho$ [28]) being 14% for *LTIA* and only 4% for *LTOA*. We conclude that $T_c$ and $H_{c2}$, which are related to the crystallite properties, are not affected by oxidation. On the other hand the critical current at high fields seems to be favoured by an increasing of grain boundaries pinning force, being improved by the presence of MgO at the grain surface despite the decreasing of connectivity. We speculate that the presence of a MgO layer of 40-50nm [8], just of the scale of the coherence length, can favour this effect.

## Conclusions

We carried out a detailed study realized by HEXRD on the phase formation and sintering in $MgB_2$ *ex-situ* P.I.T. tapes. The analysis was performed during the $MgB_2$ synthesis realized starting from $MgH_2$ and amorphous B in powder form as precursors: we focused on the effects of the reaction temperature, the oxidation process and sintering process of the tapes on the phase composition and the microstructure and tried to correlate these properties with the critical current in field in the final tapes. We have found that a solid-state reaction between the precursors occurs at quite low temperature, below 300°C, leading to the formation of $MgB_2$, while the synthesis is complete in the temperature range 700-780°C. We investigated the sintering process of the tapes in inert atmosphere and found that the cell parameters *a* and *c* for the sample synthesized at *HT* closely follow the temperature profile, while the sample synthesized at *LT* undergoes a strong rearrangement of the cell edges in the temperature region 750-900 °C due to a greater stress recovery. We also estimated the lattice rearrangement during the sintering process through the broadening of the FWHM of the 100 peak, which resulted broader for the sample cooled in oxidizing atmosphere below 400°C. From the comparison of the HEXRD patterns of samples synthesised under different atmosphere (IA and OA) we saw that the OA sample has higher MgO content, being consistent with showing better $J_c$ behaviour at higher magnetic fields.

Acknowledgements:

The authors wish to thank: "Compagnia S. Paolo", the PRIN 20082BBZ9W and the FIRB "MAST" (RBIP06M4NJ_001) for their financial support.

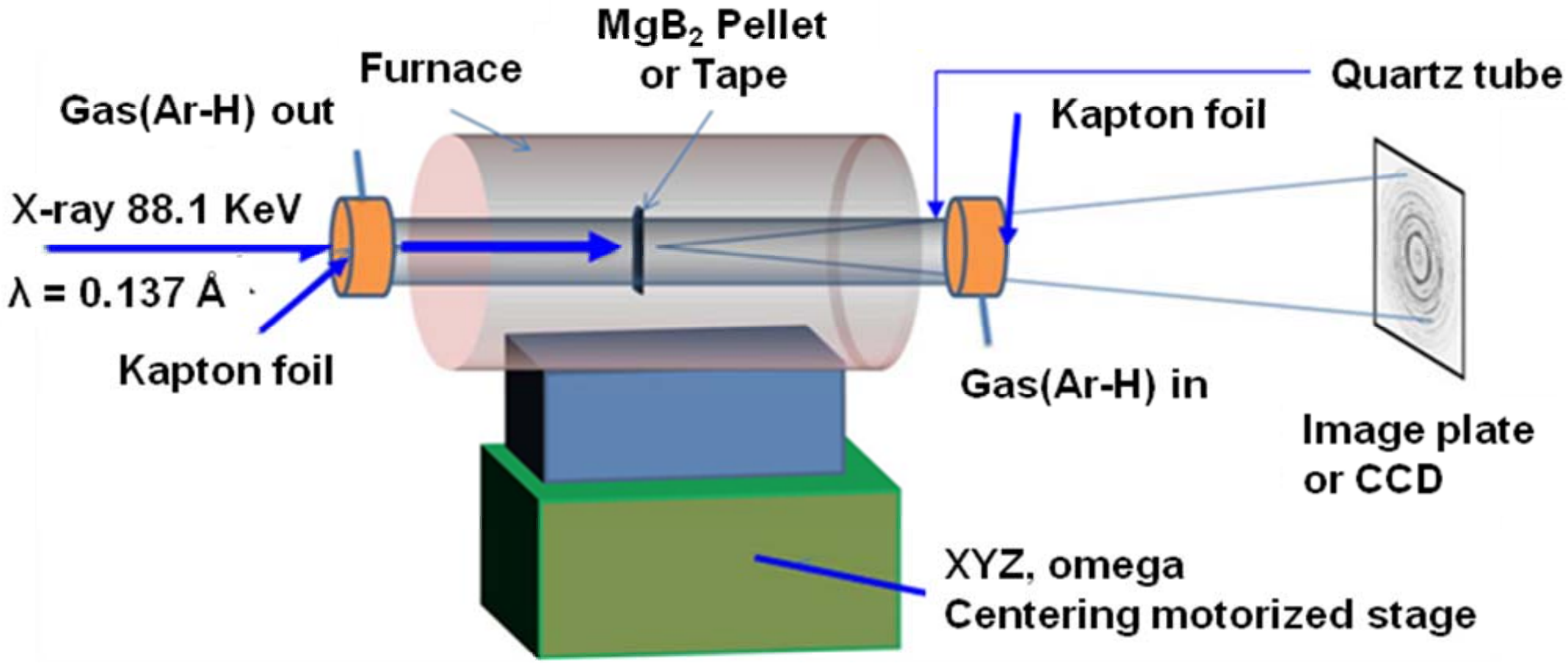

**Figure 1.** Experimental set-up for the transmission HEXRD analysis at ID11 beamline of ESRF.

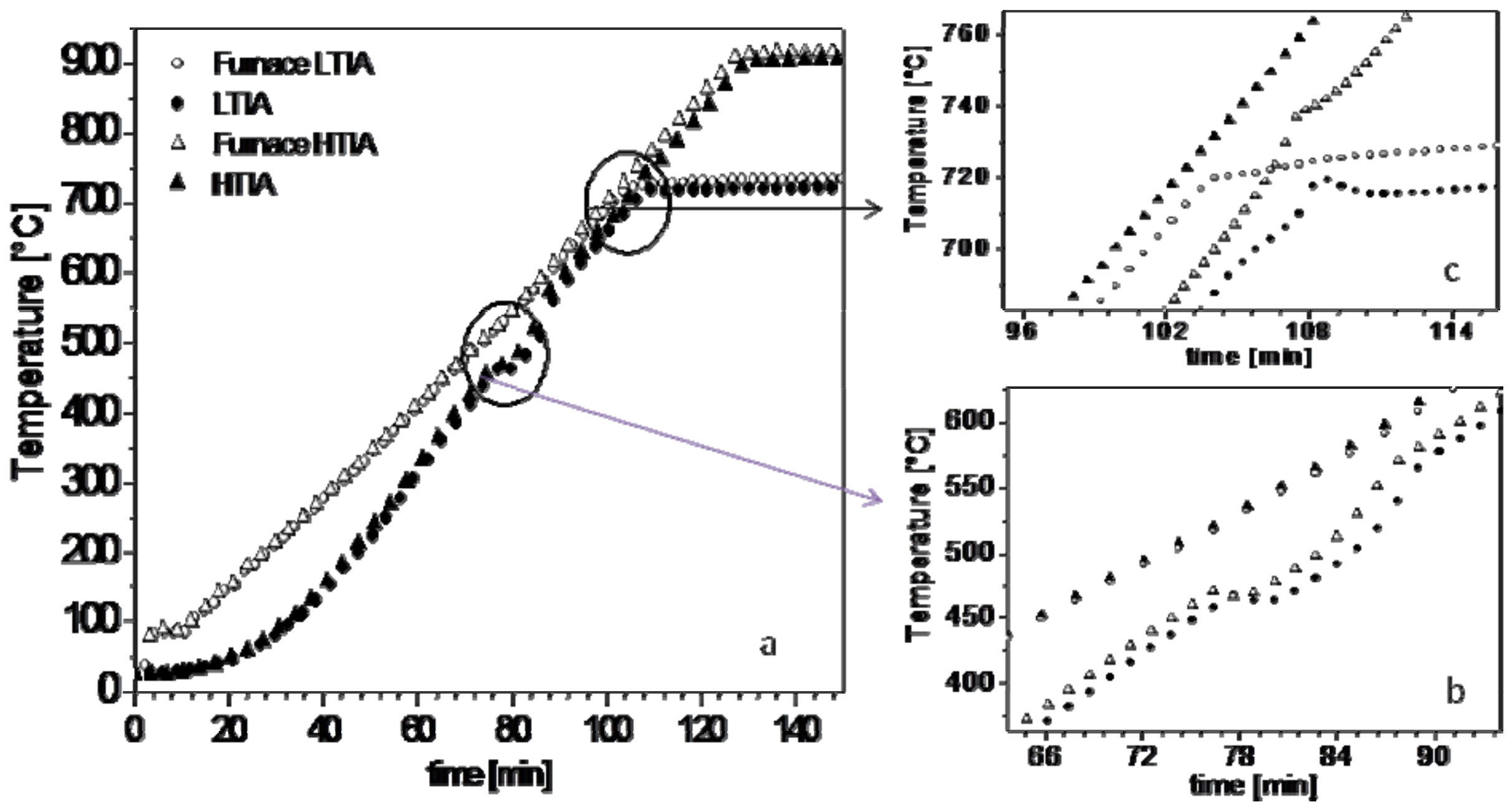

**Figure 2.** Thermal analysis of the $MgB_2$ synthesis reaction (a) temperature profile of the thermal treatments for the low and high synthesis temperature. (b) details of $MgH_2$ decomposition and of $MgB_2$ formation (c).

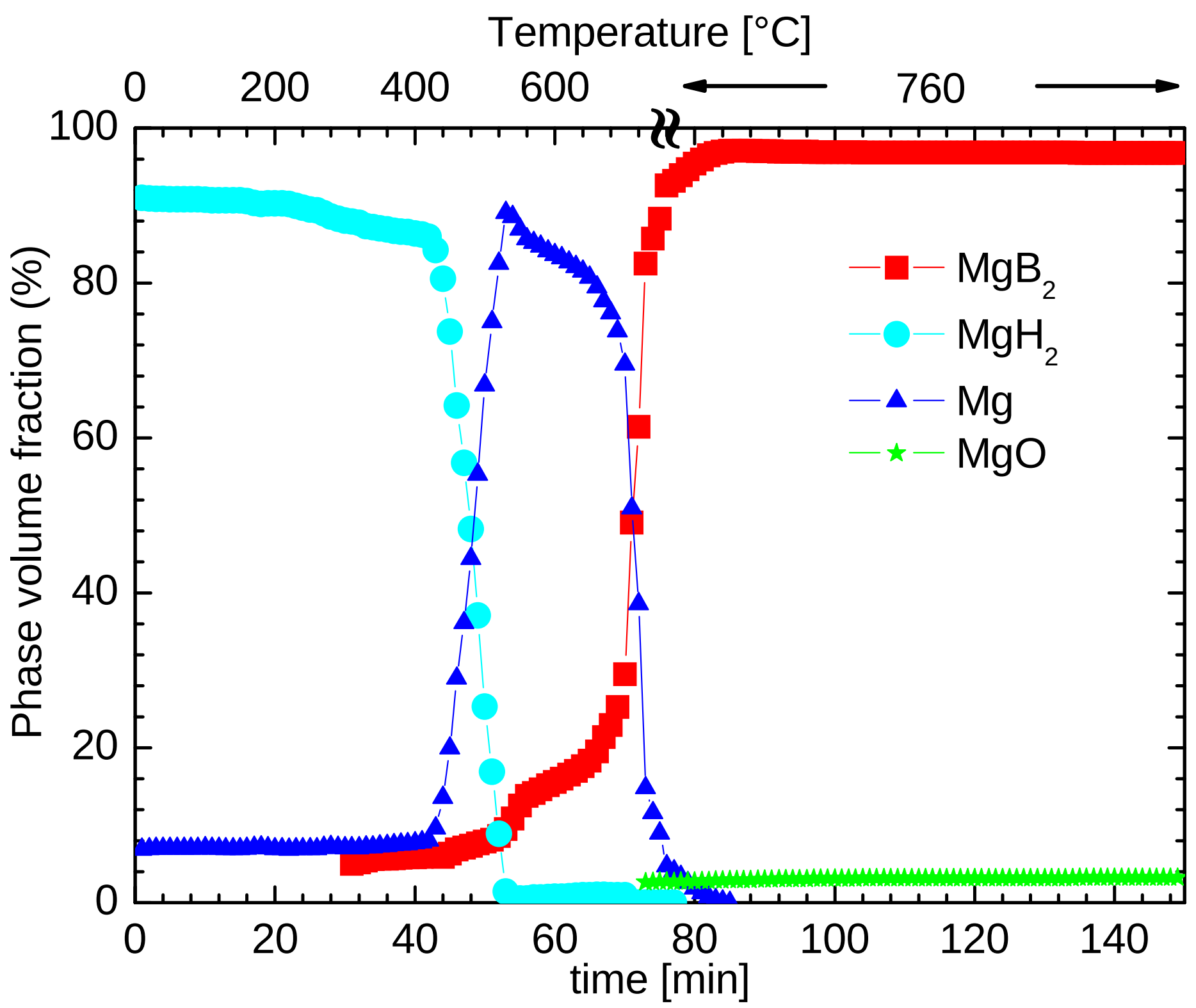

**Figure 3.** Phase volume fraction evolution as a function of the time during the synthesis of $MgB_2$ at low temperature (760°C).

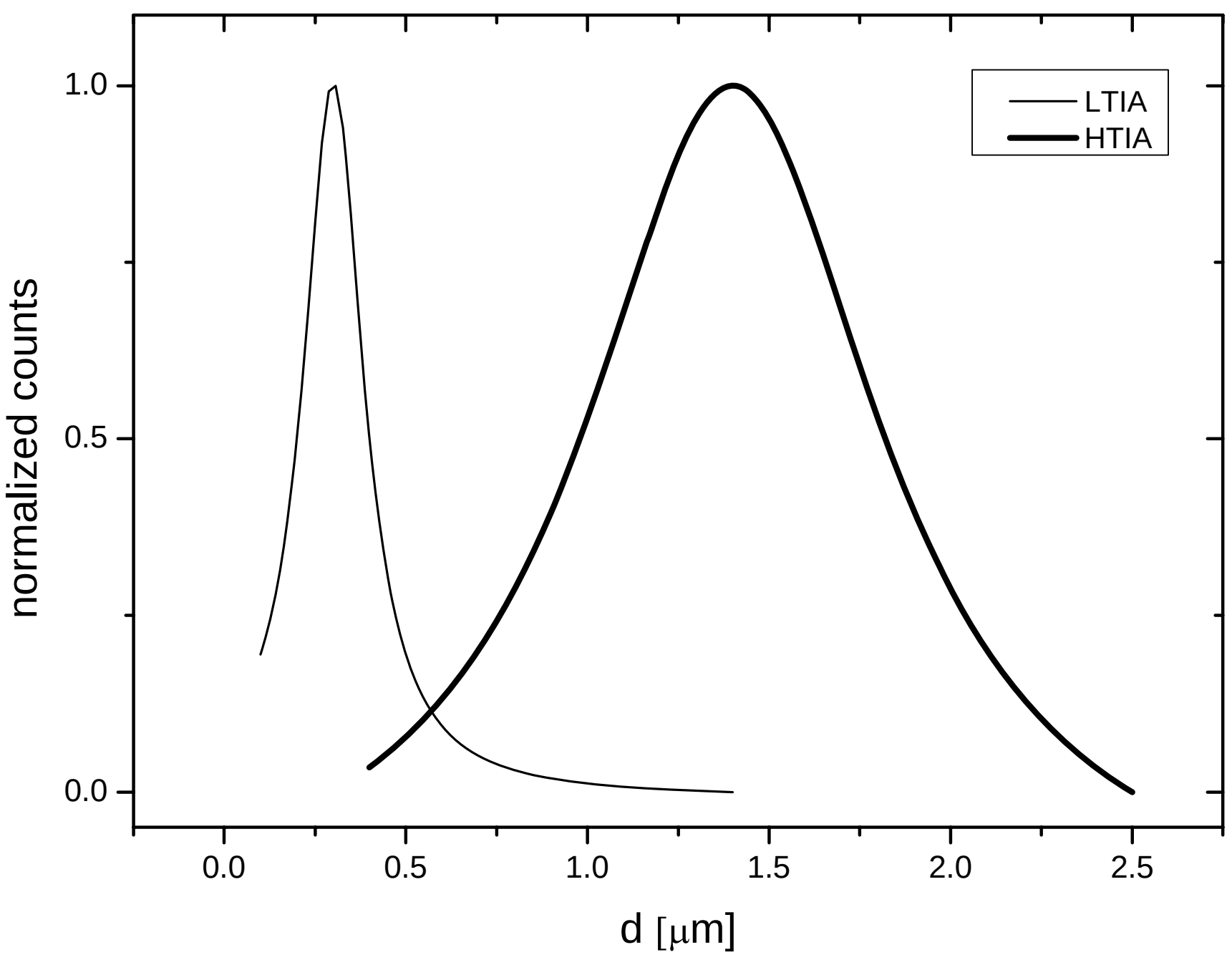

**Figure 4.** granulometry distribution fitted with a Lorentzian function for the low temperature synthesis sample (LTIA) and for the high temperature synthesis (HTIA) one.

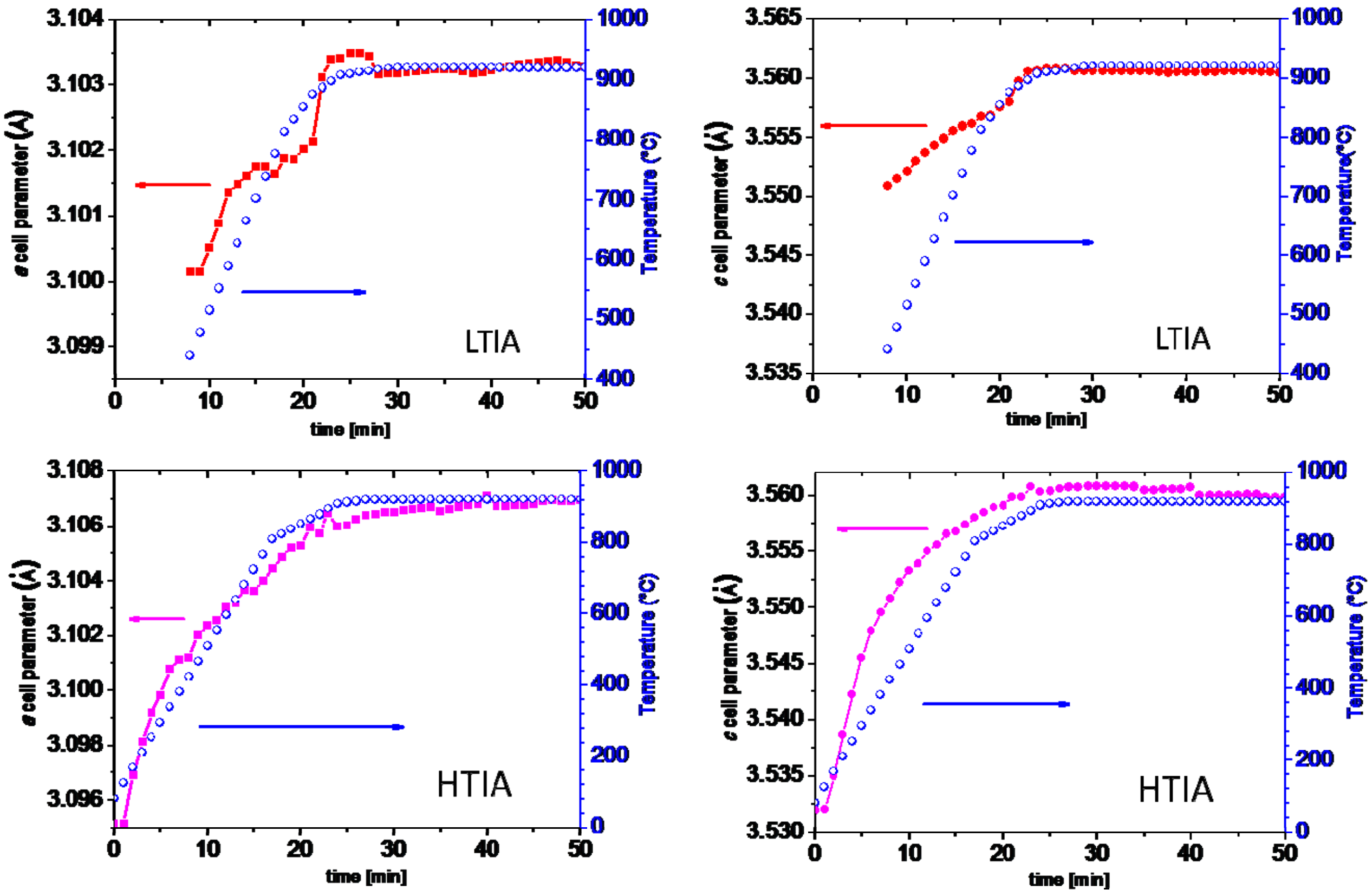

**Figure 5.** $MgB_2$ lattice parameters evolution during the sintering process for LTIA (upper panel) and HTIA (lower panel) samples.

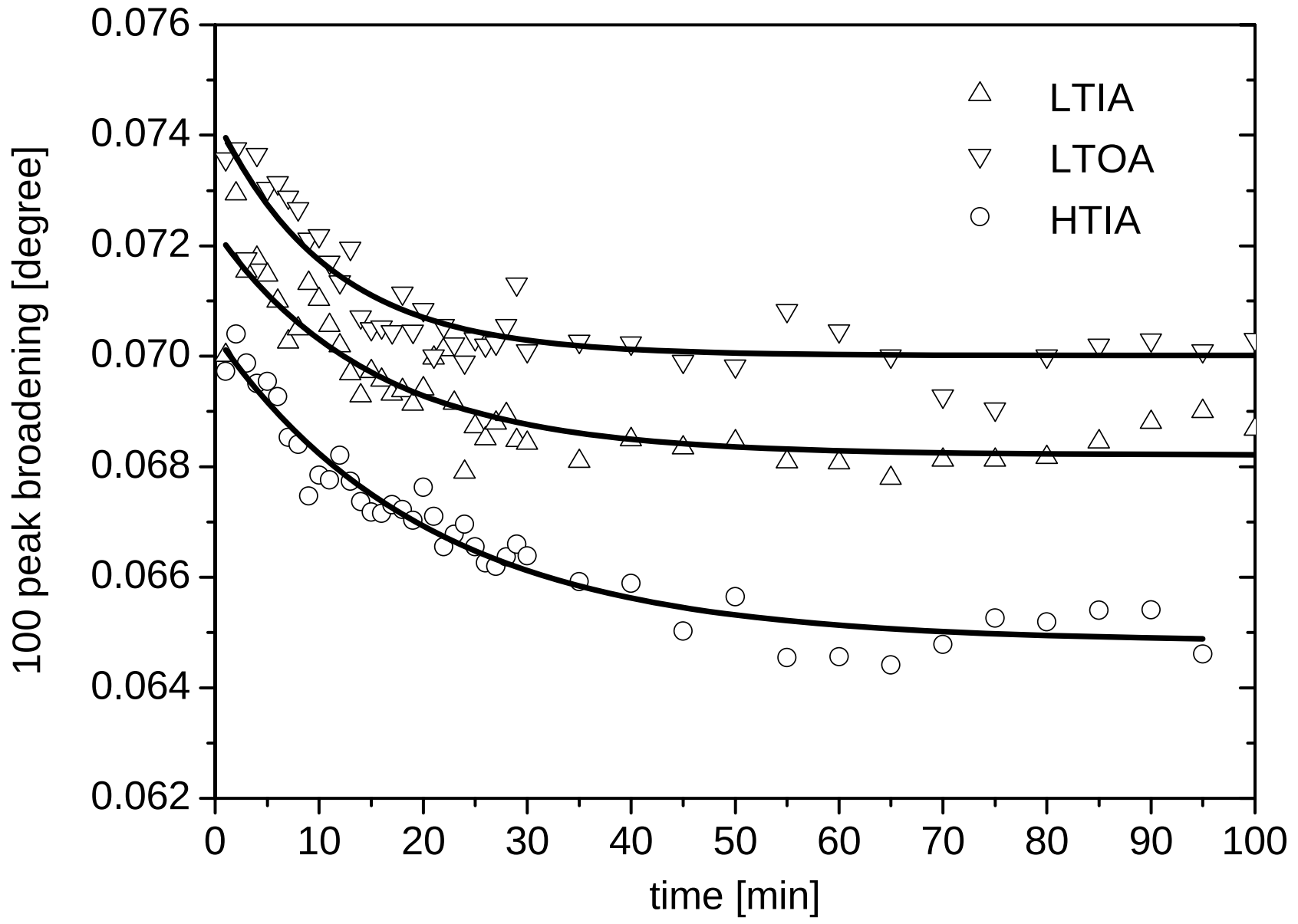

**Figure 6.** Broadening of the 100 peak of $MgB_2$ as a function of sintering time for different synthesis temperature and atmosphere. The continuous lines represent the exponential fit of each data set.

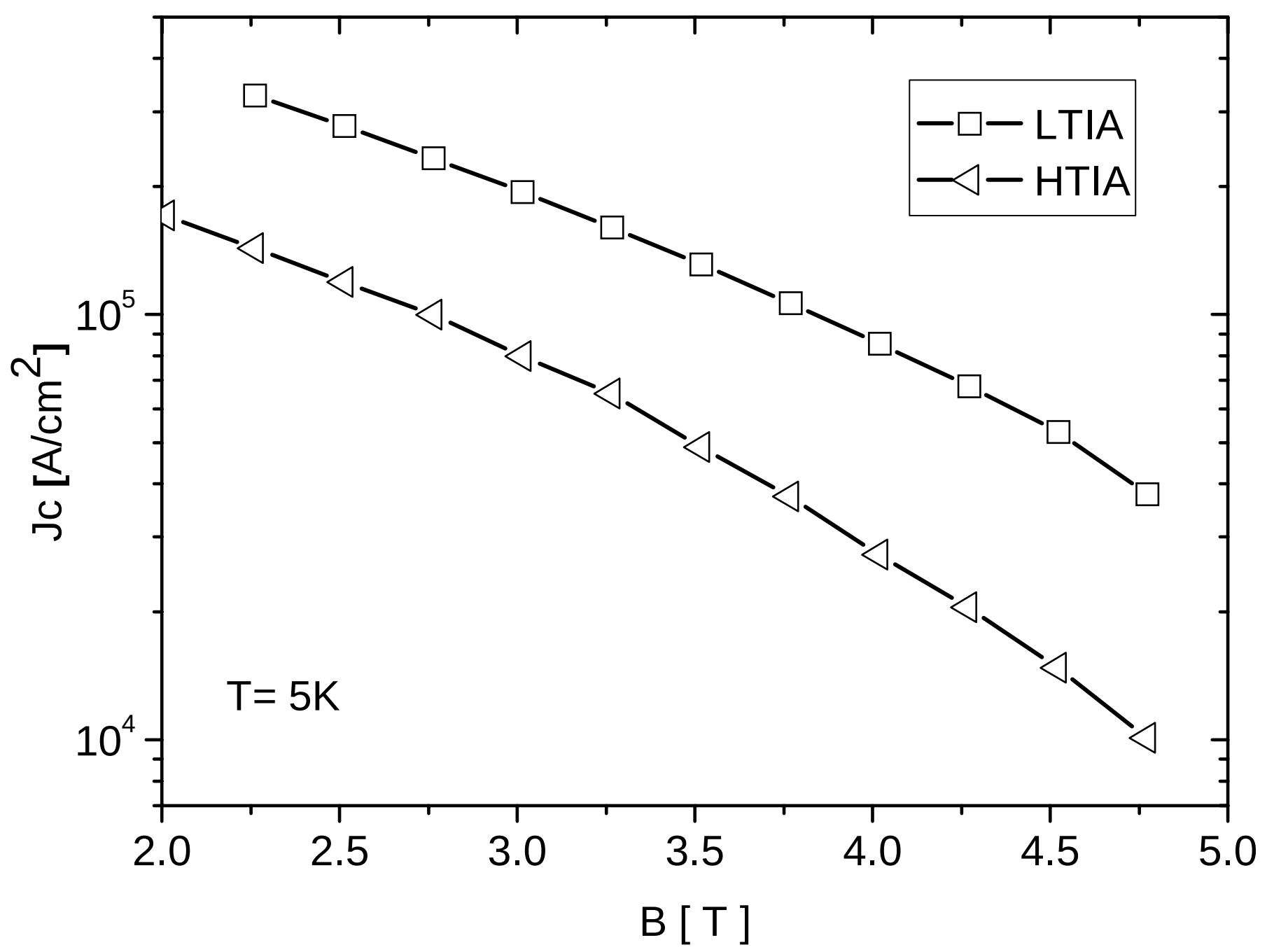

**Figure7.** $J_c$ measurements for the wires of reference [5] (synthesis temperature for LTIA = 745°C).

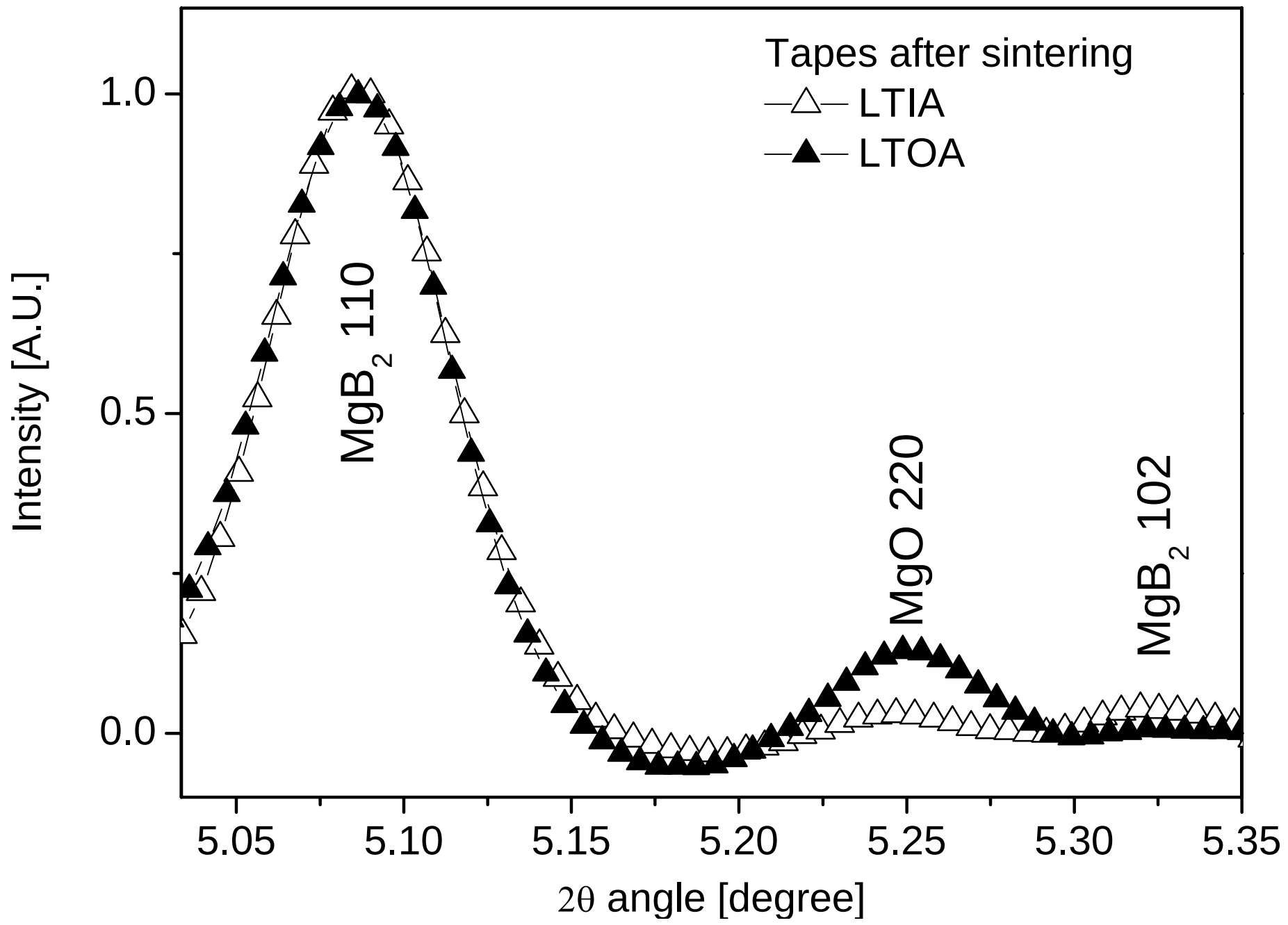

**Figure 8.** Enlarged view of the HEXRD patterns of samples synthesised under different atmosphere (IA and OA) both at low temperature (LT), the patterns have been acquired during the sintering process of respective tapes, here only the last pattern for each is reported.

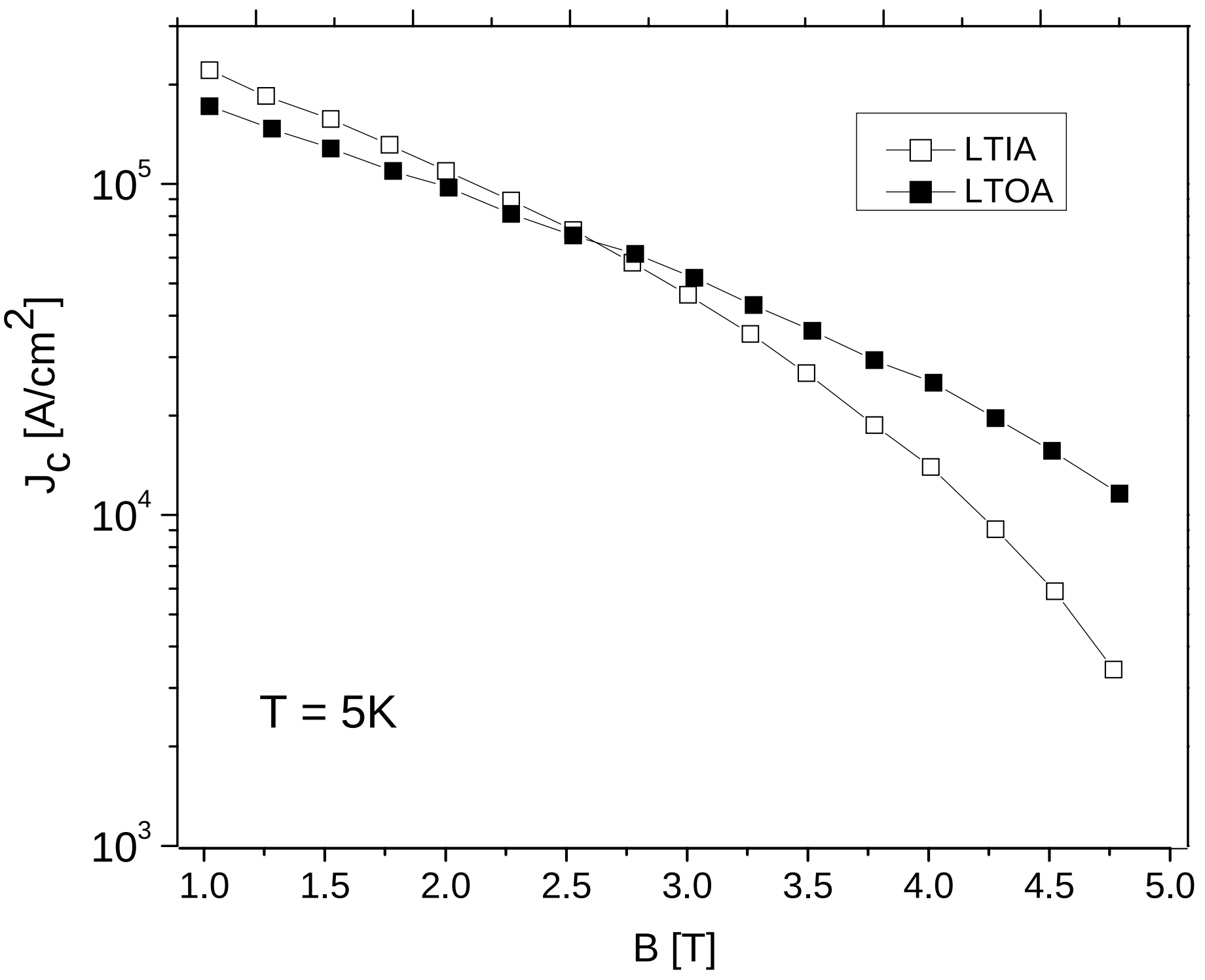

**Figure 9.** Magnetic $J_c$ of the sample synthesized at low temperature in different atmosphere reference [7].